# Dramatic reduction of dimensionality in large biochemical networks due to strong pair correlations

Michael Dworkin[1,4], Sayak Mukherjee[1,2], Ciriyam Jayaprakash[1,3] and Jayajit Das[1,2,3,5]

[1]Battelle Center for Mathematical Medicine, The Research Institute at the Nationwide Children's Hospital and Departments of [2]Pediatrics, [3]Physics, [4]Mathematics and [5]Biophysics Graduate Program, The Ohio State University, 700 Children's Drive, Columbus, OH 43205

**Abstract:**

Large multidimensionality of high-throughput datasets pertaining to cell signaling and gene regulation renders it difficult to extract mechanisms underlying the complex kinetics involving various biochemical compounds (e.g., proteins, lipids). Data-driven models often circumvent this difficulty by using pair correlations of the protein expression levels to produce a small numbers (<10) of principal components, each a linear combination of the concentrations, to successfully model how cells respond to different stimuli. However, it is not understood if this reduction is specific to a particular biological system or to nature of the stimuli used in these experiments. We study temporal changes in pair correlations described by the covariance matrix between different molecular species that evolve following deterministic mass action kinetics in large biologically relevant reaction networks and show that this dramatic reduction of dimensions (from hundreds to <5) arises from the strong correlations between different species at any time and is in sensitive of the form of the nonlinear interactions, network architecture and values of rate constants and concentrations over a wide range. We relate temporal changes in the eigenvalue spectrum of the covariance matrix to low-dimensional, local changes in directions of the trajectory embedded in much larger dimensions using elementary differential geometry. We illustrate how to extract biologically relevant insights such as identifying significant time scales and groups of correlated chemical species from our analysis. Our work provides for the first time a theoretical underpinning for the successful experimental analysis and points to way to extract mechanisms from large-scale high throughput data sets.

**Introduction**

The ability to examine living organisms on molecular, cellular and organ-level scales in high-throughput experiments offers us a detailed view of the complex nature of life [1-3]. However, the enormous amount of data makes it is difficult to extract underlying mechanisms and construct predictive mechanistic models that are often built using a small number of effective variables. This problem has been sidestepped by using data driven models endowed with predictive powers but that do not readily yield underlying mechanisms [4, 5]. A class of these data-driven models is based on the calculation of pair correlations[6, 7] of the expression levels in data sets of large number of signaling proteins at different times[1, 4, 8-12]. A startling feature of these models[1, 4, 8-11] is the presence of a very small number (often fewer than 5) of principal components, each a linear combination of a large number of variables in the data set, that can successfully describe input-output relations in the system. Even though in signaling networks activity of a signaling protein represented as a node in the network usually depends on activities of other proteins or neighboring nodes, the presence of the connectivity does not guarantee strong correlations in the activities of the proteins at different times. In the case of $n$ coupled simple harmonic oscillators, one obtains a band of frequencies for the $n$ normal modes and many modes will be necessary to describe the variations of the system in the phase space[13]. Therefore, this dramatic reduction of dimensionality in data sets indicating strong correlations between different proteins in a network remains poorly understood. Our work addresses the question of whether the decrease in the effective dimensionality in the study of two-point correlations is a general feature of chemical reaction networks that are relevant to biological systems. If this is true what biologically important results can we obtain? Answering these questions is crucial for obtaining biologically insightful mechanistic models and has inspired intense research aimed for obtaining a reduced description of biochemical networks[14-16].

Data driven models use multivariate statistical techniques such as principal component analysis (PCA) to reduce dimensionality in large-dimensional data sets. When variables in a dataset are closely related to each other or are tightly correlated pairwise, then it is possible to construct fewer numbers of new independent "super-variables", linear combinations of the original variables, that can represent the dataset. The greater is the strength of the correlation, the smaller is the number of new variables required to represent the original dataset. This is the basic principle used by PCA to construct a smaller number of new variables or principal components and the "super-variables" generate a much lower dimensional subspace where most of the variance in the data set can be captured. The presence of such lower dimensional space is evident if we are able to visualize the data points. For example, in a data set containing two variables (Fig. 1A), the data points can be represented by points in the Cartesian plane spanned by the coordinates representing the variables in the dataset. If the points in the graph lie predominantly along a particular direction indicating strong correlations between the variables, a single principal component chosen along that direction can capture more of the variability in the dataset. However, in the absence of such correlations, data points can be spread out along both directions (Fig. 1A) and will not lead to any reduction of dimensions by PCA.

We show using pair correlations in the time-dependent (or non-steady state) data obtained from numerical solution of coupled ordinary differential equations (ODEs) describing mass-action kinetics of large, biologically significant sets of biochemical reactions that the dramatic reduction (from hundreds to fewer than 3 in most cases) of the dimensionality in the system is surprisingly general. Our study demonstrates that this reduction occurs regardless of the form of nonlinear interactions in biochemical reactions, network architecture, and, persists over a wide range (about 100 fold change) of parameter values (rate constants and concentrations) for particular reaction networks. This reduction provides us with a small set of effective variables that could be useful in extracting mechanisms and identifying the key processes in complex biological systems.

Major advances in the physical sciences have involved a similar systematic thinning of degrees of freedom in complex systems when we describe phenomena on certain length and time scales[17]. Consider Euler's introduction of the moment of inertia reducing the description of the dynamics of a rigid body to a few co-ordinates instead of details of all the particles, or, the few macroscopic variables used in the description of many particle systems in thermodynamics and hydrodynamics. The key feature underlying these examples is the correlations (typically, pair) induced by the dynamics or constraints as in the case of the rigid body[17]. Our study is motivated by these considerations and investigates the generality of the dimensional reduction that results from data analysis.

**Results:**

*Linear Networks:* We first investigated whether the substantial decrease in dimensions occurs in a model system of a linear network of reactions (Figure 1B) described by linear mass action kinetics[18]. Such networks could occur for special cases of enzymatic phosphorylation and dephosphorylation of multiple sites in protein molecules[19]. The simplicity of the network architecture and the linearity of the interactions of the basic model allow us to investigate a set of progressively complex models in which we individually vary the number of molecular species and the network architecture, and introduce non-linearity into the mass-action kinetics. The kinetics of the system is given by a set of linearly coupled ODEs: $\frac{dc_i}{dt} = \sum_{j=1}^{N} M_{ij} c_j$, where $c_i$ is the concentration of the $i^{th}$ species and the non-vanishing elements of $M$ are determined by the network architecture and are linear functions of the rate constants. We use the pairwise, time-averaged, equal time covariance between concentrations ( $c_1(t), \cdots, c_N(t)$ ) of signaling proteins recorded at times $\{t_1, \ldots, t_n\}$ obtained from solving the rate equations describing the mass action kinetics of the network. The time averaging, as we show later, incorporates details regarding the shape of the kinetic trajectory of the system in the space spanned by species concentrations over a time interval in the correlation function. This is in contrast to the equilibrium equal time pair correlation functions used in physics. We obtain the covariance matrix at different times by averaging over the temporal evolution[7, 20, 21], $C_{ij}(t_n) = n^{-1} \sum_{\alpha=1}^{n} \left( c_i(t_\alpha) - \bar{c}_i \right)\left( c_j(t_\alpha) - \bar{c}_j \right)$, where, $\bar{c}_i = n^{-1} \sum_{\alpha=1}^{n} c_i(t_\alpha)$, as we vary the time interval ($t_1$ to $t_n$) of observation. The principal components, $\{\hat{p}_a\}$, constitute a set of orthogonal $N$ dimensional unit vectors), evaluated from the eigenvalue problem, $(\mathbf{C} - \lambda_a \mathbf{I})\hat{p}_a = 0$, in each time interval ($t_1$, $t_n$). The principal components provide a set of basis vectors that maximize the variance of the projection of the data points ($n$ points in $N$ dimensions) on these vectors[7, 20]. A measure of the variance captured by a particular principal component, $\hat{p}_a$, is given by the ratio, $\lambda_a / tr(\mathbf{C}) \times 100\%$, known as the *percentage explained*[20]. The above procedure is widely used in multivariate statistics literature [6, 7, 21] and is known as principal component analysis (PCA) and the principal components ( $\xi^a$ ) are often defined using the above unit eigenvectors ($\{\hat{p}_a\}$) and the data matrix ($c_{i\alpha} = c_i(t_\alpha) - \bar{c}_i$ ) as, $\xi^a_{1\alpha} = \sum_{i=1}^{n} (\hat{p}_a)^T_{1i} c_{i\alpha}$. We will refer to $\{\hat{p}_a\}$ as the principal [17]components throughout the text without any loss of generality.

Our results for the reduction of effective dimensions using principal components are based on the large percentage explained by the few components in different biochemical networks. We found that as the system is evolved in time the principal component corresponding to the largest eigenvalue captured most (~90%) of the variance in the system most of the time (Figure 1C). Only in a few short intervals of time (Figure 1C) did the variance associated with the component decrease significantly (~50%). When the top two eigenvalues were included the percentage explained rose to 90% in all the time intervals. (Supplementary Figure S1). We display our results in Figure 1B until the time when the system has evolved very close (See details in the supplementary material) to the steady state when the eigenvalues of the correlation matrix become stationary in time. Even when the rate constants and the initial concentrations were varied 100-fold over a uniform, random distribution, the largest eigenvalue of *C* captured a significant fraction of the total variance (Figure 1C). By examining the trajectories (see later) we associate the decrease in the percentage explained in short time intervals with significant changes in the direction in the trajectory of the system in the space of species concentrations; however, such changes do not occur very frequently and appears to hold regardless of the architecture and the number of species in linear networks. We provide a more quantitative connection between the change in the percentage explained and the change in the shape of the system trajectory, and how such changes in the percentage explained can be used to extract biologically significant timescales later in the paper.

To quantify the results further, we generated an ensemble of eigenvalues of the covariance matrix calculated at different times until the steady state is reached for a 100-fold variation of rate constants and initial concentrations. The distribution of the percentage explained by the largest eigenvalue calculated at all times for all trials (representing a particular choice of rate constants and initial concentrations) of a given number of species *N* is shown in Figure 1D. As *N* is increased, the distributions became wider and with peak values around 99% for $N \geq 64$. The distribution function appears to converge to a limiting function when *N* is increased further. This is also reflected in the saturation of the average percentage explained (inset, Figure 1D) as a function of *N* within the limitations of our numerical computations. When the top two eigenvalues are considered, the distributions are peaked around 100% (Figure 1E) for all the networks studied with the mean values above 95% for *N*>64 (inset, Figure 1D). This behavior can also be quantified using the metric of the minimum percentage explained in any trial and is shown in the web-supplement (Supplementary Figures S2A-B). Our results are robust with respect to variation of initial conditions or inclusion of conservation of subsets of species as shown in the Supplements. A theoretical analysis of these results is presented in the supplementary material and in the Supplementary Figure S10.

The above results show the efficacy of principal component analysis in linear networks; real biological networks are more complex and so we describe below our results for networks with more complex architecture and nonlinear rate equations.

*Variation in network architecture:* We changed the architecture of linear cascades to form random linear networks with one or more disjoint reaction modules (web supplement). Such changes did not result in any changes in the quantitative conclusions (Figure 1F) drawn from our earlier analysis. These results show that species interacting in chemical reactions evolve in time with a high degree of correlation regardless of the network topology.

We have devised a counter-example, (details in the web-supplement), a set of disjoint linear networks where all the principal components are required to capture the variation; however, this requires a strong temporal co-ordination in the initial conditions.

*__Nonlinear networks:__* We tested whether the earlier results arise from the linearity of the ODEs, by adding feedbacks between neighboring species changing both the network topology and introducing nonlinear interactions in the linear cascade. The results showing the distribution of the minimum percentage explained by the largest component in a set of trials varying rate constants and initial concentrations are displayed in Figs. 2A and 2B. The shape of the distributions appears to converge for $N \geq 64$ with a peak around 42% (Figure 2A). Once again the mean of the distributions saturates as $N$ as is increased (Figure 2B). Figure 2B shows that the largest eigenvalue is less competent in capturing the variance in the data as compared to linear kinetics. Nevertheless, when the top four components are considered the peaks of the distributions moved close to 100% with a small spread ($\sigma = .05\%$) around the peak (Supplementary Figure S5). This demonstrates that non-linearity of the ODEs does not significantly affect strong correlations between different molecular species.

*__Signaling and gene regulatory networks:__* We chose three important biological networks (i) to probe the utility of the method in capturing variations in biologically relevant signaling networks validated in experiments and (ii) to investigate the effects of biologically relevant motifs and kinetic features, such as bistability and oscillatory behavior, absent in linear networks.

First, we discuss the results for a signaling module with a positive feedback in lymphocytes involving Ras activation[22, 23]; modules with positive feedbacks are commonplace in cell signaling networks and are instrumental in producing decisive responses in these systems[22, 24]. T cell receptors interacting with antigens initiate a series of signaling events recruiting two key enzymes, SOS and Rasgrp1, to the plasma membrane that activate a G protein Ras in T cells (Supplementary Figure S6A). Activation of Ras results in priming of the MAP kinase pathways to affect cell differentiation and proliferation. A positive feedback in Ras activation is mediated by the enzyme, SOS, producing bistability in the steady state of the Ras activation kinetics arising due to a pitch-fork bifurcation[25]. Since, RasGTP produced by Rasgrp1 activates the positive feedback induced by SOS, the interplay between these two enzymes controls the time scales for robust Ras activation and the manifold of bistability in the steady states. A specific motivation to study this network is to investigate how nonlinear interactions between the enzymes affect correlations between different species and thus the number of principal components required to describe the Ras activation kinetics. We found that most of the variations (>69%) in the data as described by the covariance matrix, *C*, can be captured by the principal component corresponding to the largest eigenvalue (Figure 2C and Supplementary Figures S7, S8A) demonstrating a high degree of correlation between the molecular species at all times. Our dataset was composed of all the species concentrations in the network (web-supplement) calculated at different times until the system reached the steady state, and we varied rate constants and initial conditions to include stable and bistable steady states.

The temporal evolution of the percentage explained by the largest eigenvalue shows that only during a few short time intervals does it decrease to smaller values (< 75%); in these time intervals significant changes in species concentrations occur (Supplementary Figure S7), as we found for the linear kinetics. In this case a large change in Ras activation produces significant changes in the other species concentrations associated with RasGTP. We show this connection more directly in a later section by studying Gram determinants.

Next we investigated kinetics in the epidermal growth factor receptor (EGFR) signaling network ubiquitous in cells in humans and other mammals that plays significant role in embryonic development and tumor progression[26, 27]. Upon binding with epidermal growth factor (EGF) these receptors undergo dimerization that results in autophosphorylation of multiple amino acid residues eventually leading to activation of Ras and MAP kinase cascades that regulate cell differentiation and proliferation. This signaling network, with hundreds of proteins, has been extensively studied in experiments and in computational modeling because of its direct role in regulating growth of tumors. The signaling network involves many nonlinear interactions, such as, dimerization, multi-site phosphorylation and de-phosphorylation and enzymatic activation and de-activation of proteins.

We investigated in a smaller EGFR signaling network model[27] containing about 19 different species (Supplementary Figure S6B) and a larger model [28] involving more than 300 species (Supplementary Figure S6C). Both the models describe signaling kinetics until Ras activation and contain the key non-linear processes. The larger model contains multiple activation states of the receptors, enzymes and adaptor proteins, e.g. the EGF receptor can contain 3 activation states instead of one in the smaller model. The results of the larger model agree qualitatively with those in the smaller model for the same parameters as the smaller model. Thus, comparison of pair-wise correlations in both the models provides insights into how retaining multiple molecular states affects the correlations between different species concentrations in a large signaling network. Such scenarios are ubiquitous in cell signaling network where proteins may contain large numbers of phosphoforms, form oligomers, or interact with other proteins at multiple sites. We found that the percentage explained by the largest eigenvalue for both the small and the larger model varies similarly as the rate constants and initial concentrations are varied (Figure 2D). The distributions of the percentage explained by the largest eigenvalues are almost identical in both models. When top three eigenvalues are used, the minimum percentage explained was >88% for all the trials (Supplementary Figure S8B). The striking similarity between the large and smaller models indicate that inclusion of multiple activation states did not introduce new time scales over which the concentrations changed significantly.

The choice of the next nonlinear network (Supplementary Figure S6D), the NF-κB network, was motivated by several reasons. Unlike the other networks this system does not necessarily reach a stable fixed point. A Hopf bifurcation[25] occurs as rate constants are varied separating fixed point behavior from sustained periodic oscillations with decaying oscillations in the vicinity of the bifurcation. This behavior has been observed experimentally and an experimentally validated model reported in Refs.[29, 30] makes this suitable for analysis. There have also been theoretical attempts to describe the model with fewer number of components[27]. Furthermore, NF-κB is a ubiquitous, highly stimulus-sensitive protein that is a pleiotropic regulator of gene induction controlling a wide range of functional behavior. The temporal behavior of the nuclear and cytoplasmic NF-κB concentrations is dynamic and determined by a delicate balance between different feedback mechanisms operating on different time scales. Thus this network provides a good testing ground for the questions we address in our work. We vary the parameters in the model to yield oscillatory, as well as, non-oscillatory trajectories. In Figure 2E we show the time evolution of the percentage explained by the top principal component. For the trajectories that show sustained oscillations or damped oscillations with slow amplitude decays, the percentage explained by the top component also oscillates with similar frequencies. The lowest values in the percentage explained for such trajectories hover around 50% while the top two principal components capture, more than 90% variation (Figure 2F). This behavior is similar to the kinetics of the principal components in a two variable system where each variable oscillates with the same frequency. Even in this multi-dimensional kinetics oscillations are effectively two dimensional. When both oscillatory and non-oscillatory trajectories are included, Figure 2G shows that the percentage explained by the largest eigenvalue is peaked around ~50% with values as low as 30% possible. When the top two eigenvalues are used, the peak in the percentage explained data moves to around 95%. For further details see Figure S8C. While many components show sustained or decaying oscillatory behavior these are not independent and are enslaved to the nuclear concentration of NF-κB. It is the correlated motion, orchestrated by a small number of significant species that underlies the occurrence of dimensional reduction. This model also points to one mechanism by which dimensional reduction might fail to occur: if the motion occurs on a high-dimensional torus as in the case of many oscillators or is chaotic[25].

*Gram determinants:* The constraints imposed by the fact that the trajectories are solutions to smooth, autonomous differential equations underlie the success of principal component analysis in providing a much lower dimensional effective description of the kinetics: the solution to the rate equations, c(t), is a differentiable curve embedded in *N*-dimensional space that is locally one-dimensional.

In this section we use elementary differential geometry to provide both an intuitive and a quantitative understanding of the mathematical reasons that underlie the remarkable reduction in the effective dimensionality. We examine the geometry of the solution of the ODEs in the space spanned by the species concentrations locally in time in contrast to the principal component analysis computations that average over time. Specifically, we examine the number of dimensions needed to describe the curve at a given instant of time, as a function of time. Our expectations for the success of principal component analysis are based on the geometry of three-dimensional curves, the extrinsic curvature that measures deviation from linearity and torsion that captures how the curve twists in three dimensions.

If the torsion of a regular curve with non-vanishing curvature vanishes, the curve lies in a plane. Thus in higher dimensions the curve has to twist rapidly in *r* different dimensions for the curve to need *r* embedding dimensions locally. Armed with this intuition we have used the Gramian matrix defined as a symmetric, $r \times r$ matrix whose elements are given by the inner products of a set of *r* vectors in *N*-dimensional space to study the geometry. The determinant of an *r*-dimensional Gramian matrix is the volume of the r-dimensional paralleotope formed by the *N*-dimensional vectors and is non-vanishing only if the *r* vectors are linearly independent and enclose an *r*-dimensional volume[31].

The solution to the ODEs obtained as a function of time can be written as an *N*-dimensional vector $\vec{c}(s)$ where *s* is the arc length. We form the Gramian matrix using the first *r* derivatives of $\vec{c}(s)$ with respect to the arc length. (See Web supplement for details.) If the determinant formed by the first *r* derivatives is non-vanishing the curve is *r*-dimensional. As an example, the results for the Ras activation network are displayed in Figure 3. The value of the Gram determinant for just $\{\vec{c}'(s), \vec{c}''(s)\}$ is shown as a function of time along with the performance of the largest eigenvalue of the covariance matrix as a function of time. We focus on the behavior around t = 200. A sharp change in the curve describing the dynamical evolution of the concentrations is reflected in a peak of the Gram determinant that precedes a decline in the percentage explained by the first component that occurs over almost 100 time units. Higher dimensional Gram determinants were found to be negligible showing that the curve is locally no more than two-dimensional providing a picture of why the top two principal components explain most of the variance.

Since the covariance matrix integrates the information about the pair correlations over time, local increases in twisting are smoothed out by time averaging and their effect on the required number of principal components is suppressed. This observation is borne out further by the results shown in the Supplementary section for the other models we have studied. This allows us to identify another mechanism by which more dimensions may be needed in principal component analysis: if changes in the trajectory occur along different dimensions in succession so that their effects add before they can decay in time. Our detailed study shows few rapid changes occur in the Gram determinants and demonstrates that the ability of the few principal components to capture a large fraction of the total variance is not simply due to washing out of rapid changes in the trajectory by time averaging. It is difficult to visualize curves in multi-dimensional space and the Gram determinant provides a convenient way of following the generalized curvatures in *in silico* models to identify emergent time scales in nonlinear systems.

**Extraction of significant timescales and correlations:** We illustrate how biologically useful, mechanistic insights can be extracted from our analysis based on the result that a few principal components dominate the covariance matrix. We show how biologically significant time scales can be obtained by studying the kinetics of the percentage explained by the top eigenvalues of the covariance matrix and how the composition of the corresponding eigenvectors can help identify correlated subsets of species that dominate the kinetics in different time intervals. These subsets could also help in building a coarse grained description of the system kinetics.

We associate changes in the temporal behavior of the eigenvalue spectrum of the covariance matrix with biologically significant time scales: these variations indicate significant changes in the direction of the system trajectory as confirmed by Gram determinant analysis. We can associate these in turn with the induction of new kinetic processes occurring on very different time scales. We demonstrate this by using *in-silico* signaling kinetics of the Ras activation network described earlier and high throughput experimental data available for the response of HT-29 human colon adenocarcinoma cells to various cytokines[8].

In the Ras signaling kinetics, the percentage explained by the largest eigenvalue of the covariance matrix, $f_{max}$, decreases in a short time interval (See Figure 3A). This can be connected to robust production of RasGTP and therefore, provides an estimate of the time scale of Ras activation. As we have shown in the previous section, this drop in $f_{max}$ occurs due to a change in direction in the trajectory of the system in the space of species concentrations. Consider the time period before robust Ras activation. The composition of the normalized eigenvectors associated with the largest eigenvalue shows that, RasGDP, complexes of RasGDP with its catalyzing enzymes (e.g., RasGTP or RasGDP bound to the allosteric sites of SOS) dominate the eigenvector (Fig. 4A). This shows that the enzymatic reactions producing active Ras with large catalytic rates act in a highly correlated manner and significantly control the kinetics compared to the other reactions in this time peroiod. After Ras is robustly activated this eigenvector is then mainly composed of RasGTP, RasGTP bound to the allosteric site of SOS, and RasGDP; the low percentage of the substrate enzyme complexes show that binding unbinding reactions of RasGTP with SOS take precedence over the enzymatic reactions producing Ras in controlling the system kinetics (Fig. 4B), as most of the Ras has been already activated by this time.

Next we used available data from high-throughput experiments to demonstrate how time scales and correlations can be extracted from experimental data sets[8]. We studied kinetics of the eigenvalues of the covariance matrix calculated from the activation profiles of 19 different proteins occurred due to stimulation of TNF, EGFR and Insulin receptors on HT-29 human colon adenocarcinoma cells by combinations of multiple cytokines (such as, TNF, EGF, Insulin) which are also the cognate ligands for the above receptors[8]. We found, that for every cytokine treatment (combination) reported in Ref. [8], the top three eigenvalues capture more than 71% of the variation in the data for all times. This is in agreement with our general conclusion that few principal components can capture a significant proportion of the variation in chemical reaction networks. However, in contrast to most of the networks we studied in the previous section where the top three principal components accounted for around 90% of the variance, in the experimental data set a significantly lower fraction was captured by the top three components. We have investigated various reasons for this difference: (i) variability in the experimental measurements, (ii) sparse sampling of molecular species the signaling network, (iii) using fold change in expression levels instead of absolute concentrations to characterize activation as done in the experimental analysis, (iv) measurements at much larger time intervals than that used in our *in-silico* studies, or, (v) the set of interacting signaling networks stimulated by cytokines and EFG could contain a specific set of non-linear interactions that we did not probe in our *in-silico* networks. We used the smaller EGFR signaling network that we studied in the previous section to evaluate the effect of the factors (i)-(iv). Based on our simulations (Fig. S11 in the supplementary material) we found that introducing these effects could reduce the percentage explained by the top eigenvectors, but the reduction is not as significant as that observed in the experimental dataset. On the other hand, our results for the network with non-linear interactions with linear architecture (Fig. 2D) showed that the top few eigenvalues in this particular network are the least ineffective in capturing variances (e.g., minimum variance captured by the top component and the top four components are 23% and 62%, respectively) compared to the other nonlinear networks we studied. This suggests that the signaling networks probed by Gaudet et al could contain nonlinear elements that contribute to the lower percentage explained by the principal components. We found that, for cytokine treatments, the kinetics of the percentage explained by the largest eigenvalue, $f_{max}$ can be used to extract time scales that are related to activation of important proteins. For example, when the cells were stimulated by 100 ng/ml TNF and 100 ng/ml EGF, $f_{max}$ drops to a lower value for a short time interval at 15 mins (Fig. 4C-D), which is related to the timescales of large activations in MK2 and JNK, two of the strong markers of cell apoptosis. The composition of the normalized eigenvector corresponding to this eigenvalue show that MK2, JNK and pIRS1-Tyr896 (phosphoform of the tyrosine residue associated with the insulin receptor) concentrations dominate the eigenvector and remain strongly correlated throughout the kinetics[32]. It shows even though the insulin receptor was not activated directly in this cytokine treatment the tyrosine residue attached to the insulin receptor is activated throughout the kinetics due to cross-talk between the TNF and EGF receptor signaling networks. This is consistent with the similar conclusion made by the authors of Ref.[8] using principal component analysis based on data neglecting the temporal variation of the components.

**Discussion**

In this paper we have studied biologically relevant linear and non-linear biochemical reaction networks containing large number of variables numerically, and shown that a substantial reduction in the number of effective variables required to capture the time variations and covariations of concentrations is possible. A similar reduction had been observed earlier in the analysis of experimental data and our study provides a theoretical basis for their success. The method allows us to identify subsets of variables that influence the dynamics on different time scales and to extract biologically significant time scales from *in-silico* and experimental data, thus contributing to mechanistic insights.

We have shown that time-averaged pairwise correlations (analyzed using the widely used method of principal component analysis)[7, 21] can be used to construct a handful of effective variables (principal components) that capture most of the variations in the time evolution of linear and non-linear biochemical reaction networks containing large number of variables with biologically important motifs and dynamical features. We showed with extensive calculations that this large reduction in the number of effective dimensions arises from the tight correlation between different variables and is not sensitive to the values of the rate constants, initial concentrations of different species and the network architecture and thus appears to be a common feature of biochemical reaction networks. We have presented a local geometric analysis that makes our results intuitive. This work provides a mechanistic basis for the success of the analysis of the experimental data (1,4,9).

The systems we have studied evolve to low-dimensional attractors such as fixed points (zero dimension) or limit cycles (few dimensions) at large times; while this does seem to play a role, we emphasize that the reduction occurs over the entire transient part of the trajectory even far from the attractor. We have checked that for a chemical reaction network that evolves to a strange attractor (chaos), the number of principal components needed to explain a large fraction of the variations can be equal to or larger than the dimension of the strange attractor. Such behavior is not common in cell signaling or gene regulatory networks. Our results show, for the first time to our knowledge, how ubiquitous and robust the reduction in dimensionality due to strong pair correlations is.

We have shown how it is possible to extract biologically important time scales using the kinetics of the top eigenvalues of the covariance matrix. The kinetics of the corresponding eigenvectors can help identify a small set of reactions that significantly change the system kinetics in a time interval. Such mechanistic insights could lay the foundation for constructing coarse-grained effective kinetics from large dimensional high throughput datasets, for example, by using the kinetics of the smooth trajectory in *N*-dimensions described by the generalized Frenet-Serret formulae [31]. We have illustrated our method *in silico* on cases in which the networks are known; in the experiments of Gaudet et al. many of the reactions are believed known. Investigating the success and drawbacks of this method to extract mechanistic insights from high dimensional data setsfor cell signaling networks where interactions between the signaling components are mostly unknown is important. We are currently working on such investigations.

We point out that linear methods such as the principal component analysis could fail to find local structures associated with faster changes in timescales that do not produce significant changes in species concentrations. In such cases using non-linear dimensionality reducing methods, such as local linear embedding [33], or methods that consider higher than pair-wise correlations [34] may help in finding reduced variables and associated time scales in the system.

In summary we have shown that the kinetics of a variety of large biochemical networks remains confined to a manifold with much smaller number of effective dimensions. This can be useful in deriving biological insights from large networks and may help in constructing predictive coarse-grained models.

**Method:** We use the rule based modeling software, BioNetGen[28], to generate time courses for the species kinetics for different networks. This software produces a set of ODEs corresponding to the mass-action kinetics describing biochemical reactions in the networks and solves them numerically using the CVODE solver. The smallest time intervals for calculating the species concentrations and the time the system takes to reach the steady state concentrations were chosen using the largest and smallest time scales estimated from the rate constants and initial concentrations in a reaction network. The details are in the web supplement and the Supplementary Table S1. We use the built in random number generator in Perl to generate uniformly distributed random values of rate constants and initial concentrations. We verify that the multidimensional space spanned by the rate constants and initial concentrations is sampled evenly in our simulations by comparing our results by using the Latin Hypercube sampling method (Supplementary Figure S9). LAPACK routines were used to calculated eigenvalues and eigenvectors of the covariance matrix. We did not perform any preprocessing on the concentration data that were generated by the ODE solver. In PCA based data driven models, bias toward variables with larger variations the independent variables (e.g., fold change of activation at different times) is removed by scaling the all the variables so that each variable has unit variance[8]. We did not perform such scaling because we aimed to extract biologically significant time scales which are often associated with large variations in concentrations of few activated species (e.g., Ras activation), moreover, we found that such variance scaling does not significantly change the ability of the first few principal components to capture a large amount of variation in the original dataset (Supplementary Figure S12). List of the number of species and the number of time points used in our simulations for the biochemical networks is provided in Table II in the web supplement. Further details regarding the simulation method are in the web supplement.

**ACKNOWLEDGMENTS.** S.M., M.D., and J.D. thank the Research Institute at the Nationwide Children's Hospital for funding. We also thank a grant from the Ohio Supercomputer Center (OSC) for partial support. M.D. is also supported by the Pelotonia Fellowship, OSU undergraduate research scholarship, and COF bioinformatics scholarship. One of us (C. J.) acknowledges support through contract HHSN272201000054C of NIAID. We thank James Faeder for help with BioNetGen and Will Ray and William Valentine-Cooper for discussions. We thank the anonymous referee (#2) for useful suggestions.

**Figure 1 Strong pair-wise correlations in linear networks. (A)** Data points show concentrations of species $C_1$ and $C_2$ participating in a chemical reaction, $C_1 \rightleftharpoons C_2 \rightleftharpoons C_3$, at different time points. For a set of reaction rates, concentrations of both the species could spread out significantly more in a particular direction (black points) and the principal component PC1 captures most of the variation in the dataset. In contrast, for another set of rate constants, when the points (green) do not lie predominantly in a particular direction, both the principal components, PC1 and PC2, are required to capture the variability in the dataset, and no dimensional reduction is achieved. **(B)** A linear reaction network of eight species (N=8). **(C)** Temporal evolution of the percentage explained by the largest eigenvalue for N=64 until the steady state is reached for 100 different trials (denoted as trial index in the y axis). The kinetic rates and initial concentrations were uniformly randomly distributed between 0.1 and 10. Each trial started with a non-zero concentration only for the last species ($c_{64} \neq 0$). The time shown along the x axis is divided by the time the system (different for each trial) takes to reach the steady state to show the relative changes in the % explained for different trials. **(D)** Distribution of the percentage explained by the largest eigenvalue calculated at different time points until the system reaches the steady state for 10,000 different trials for networks containing 8 to 512 species. The y axis shows the percentage of samples (each sample corresponds to a value of the percentage explained by the top eigenvalue at a time for a trial) that lie in a bin-size of 0.5%. The inset shows the average percentage explained by the top eigenvalue. **(E)** Distributions of the percentage explained by the top two eigenvalues for the same cases as in (D). The averages of the distributions are shown in the inset of (D). **(F)** Distribution of the percentage explained by the largest eigenvalue calculated at different times for linear networks with random connections (details in the web-supplement) for *N*= 64 to 512 for 10,000 trials. The inset shows the average percentage explained by the top and the top two eigenvalues.

**Figure 2 Large reduction in dimensionality in nonlinear and biologically relevant networks. (A)** Distribution of the minimum percentage explained in networks with linear topology but with nonlinear mass-action kinetics in the entire span of the temporal

evolution by the largest eigenvalue for 10,000 trials and the same variation of reaction parameters as in Figure 1. All the trials were started with a non-zero concentration only for the last. **(B)** Average minimum percentage explained by the top and the top two eigenvalues calculated from the distributions in (A). **(C)** Distributions of the minimum percentage explained by the largest eigenvalue for the Ras activation network for 10,000 trials as the 23 kinetic rates and 20 initial concentrations were varied around the base parameters (web supplement) by up to 50% from a uniform random distribution. The distributions that correspond to bistable or stable steady states are shown in red and green, respectively. **(D)** Distributions of the minimum percentage explained by the largest eigenvalue for the smaller (Ref. [27]) and the larger (Ref. [28]) network describing EGFR signaling for 10,000 trials. In each trial the kinetic rates and the initial concentrations were varied by multiplying the base values (web supplement) by a random number distributed uniformly between 0.1 and 10. **(E)** Temporal evolution of the largest eigenvalue for the NF-κB signaling network with 26 species and 38 kinetic rates. Different trials (indexed in the y axis) correspond to random reaction parameters drawn for a uniform distribution ( from 0.6 to 1.4) are shown. The time in the x axis is scaled by the time the system takes to the reach steady state or with a time span that includes 50 oscillations as appropriate. The oscillations in the species concentrations produced damped oscillations in the percentage explained (trials 50-100). **(F)** Kinetics of the top two eigenvalues for the same trials as in (E). **(G)** Distributions of the minimum percentage explained by the largest eigenvalues for the NF-κB network for 10,000 trials corresponding to trajectories with sustained oscillations (in red) and damped or no-oscillations (in green) separately.

**Figure 3 Gram determinant for the Ras activation network. (A)** The Gram determinant in the space of the first two derivatives of the trajectory c(s) with respect to the arc length s  in blue as a function of time. It has a maximum around t=200 and is small for t> 300. In red is the percentage explained by the top eigenvalue of the correlation matrix whose performance declines as the Gram determinant value increases; since the correlation matrix is a time-integrated quantity the effect of a local increase in the Gram determinant persists for a while. The value of the Gram determinant has been multiplied (by 56300) and shifted (by 90) in order to scale the maximum value of the Gram determinant to 100% for a better visual comparison with the percentage explained data. **(B)** The trajectory projected on to a three-dimensional subspace of the network. The orange dot corresponds to the time at which the Gram determinant is a maximum and the figure clearly shows that the trajectory is locally two-dimensional as reflected in the rising Gram determinant. Details regarding the correspondence between the Gram determinant, percentage explained by the top eigenvalue, and the corresponding eigenvectors at different time points are shown in a movie available at http://planetx.nationwidechildrens.org/~jayajit/pairwise_correlations/ .

**Figure 4 Extraction of biological insights from pair-correlations. (A)** Shows the composition of the eigenvector corresponding to the largest eigenvalue for the Ras activation network at a time (T=150) before  robust Ras activation at $T \approx 200$ . The square of the fraction occupied by a particular species is shown along the x axis. The parameters used for the signaling network is the same as in Fig. 3 where the kinetics of the percentage explained by the largest eigenvalue is shown. The entire time course of the eigenvector is shown in a movie available at,
http://planetx.nationwidechildrens.org/~jayajit/pairwise_correlations/ . **(B)** The composition of the eigenvector as described in (A) is shown at a later time (T=350) after Ras is robustly activated. **(C)** Kinetics of the percentage explained by the largest eigenvalue shown at different time points for HT-29 human colon adenocarcinoma cells stimulated by TNF and EGF. We used the data available at http://www.cdpcenter.org/resources/data/gaudet-et-al-2005/ for the calculation and the time points shown on the graph correspond to the times (in minutes), 0,5,15,30,60,90,120,240,480,720,960,1200,1440, of the measurements as reported by Ref.[8]. The temporal evolution of the eigenvector corresponding to this eigenvalue is shown in a movie available at
http://planetx.nationwidechildrens.org/~jayajit/pairwise_correlations/ .
**(D)** Shows the kinetics of the different species for the system described in **(C)**. The time scale associated with the decrease in the percentage explained by the largest eigenvalue corresponds to the timescales of activation of the species shown in colors.

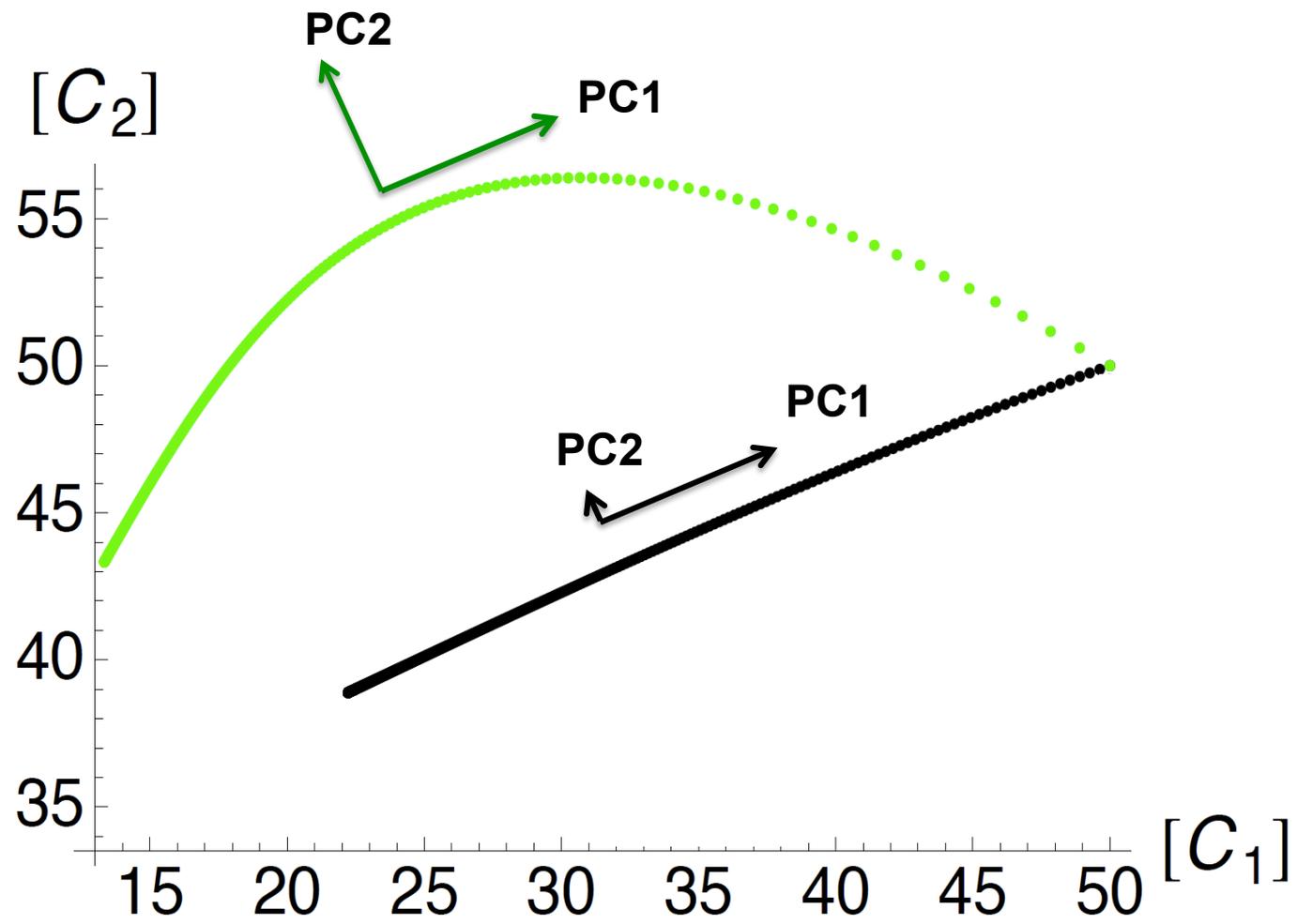

(A)

Fig. 1

(B)
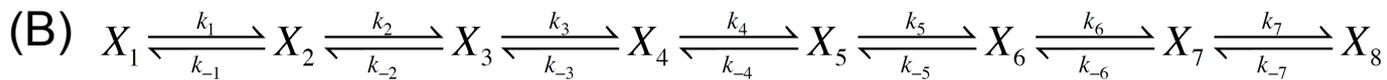

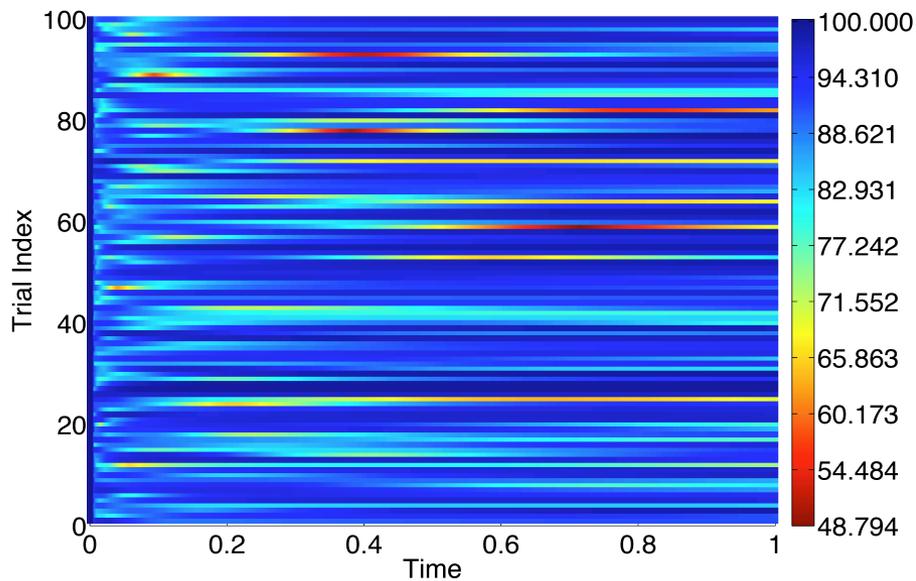

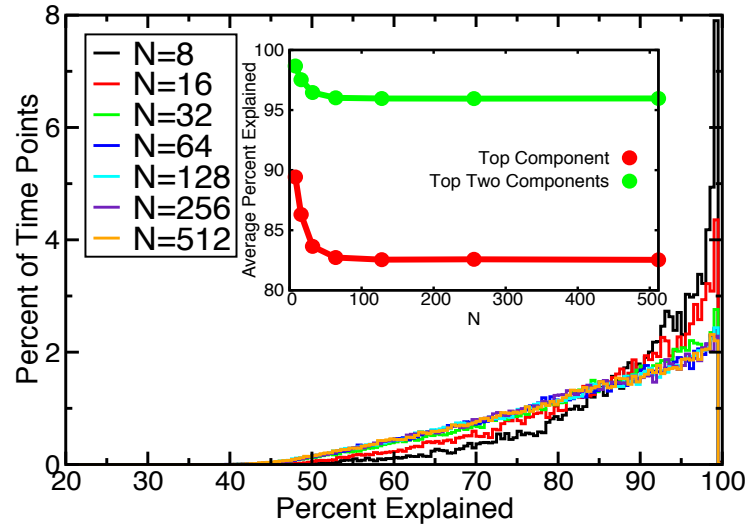

(C)

(D)

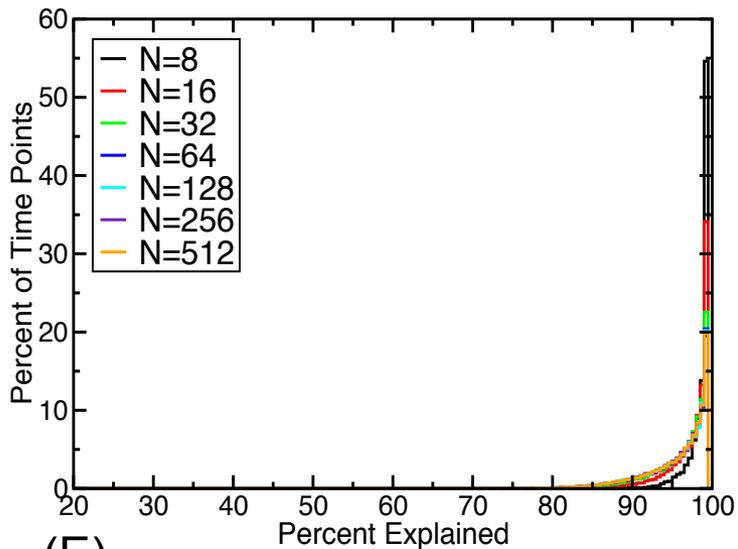

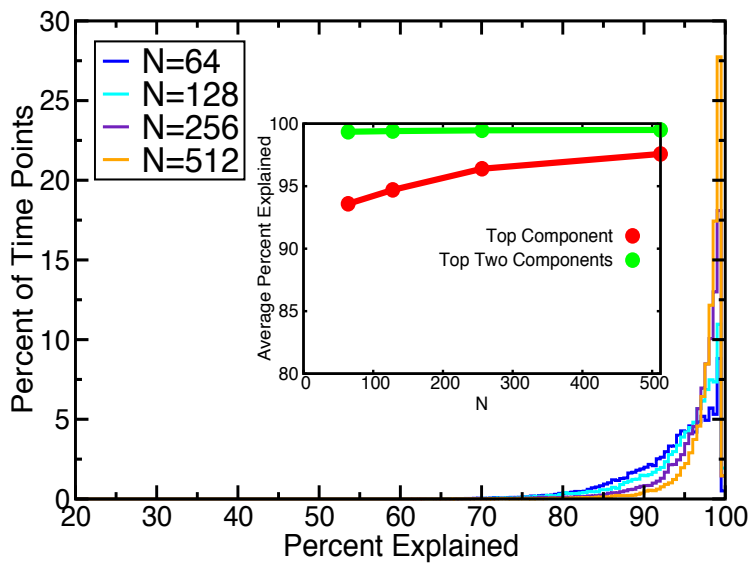

(E)

(F)

**Fig. 1**

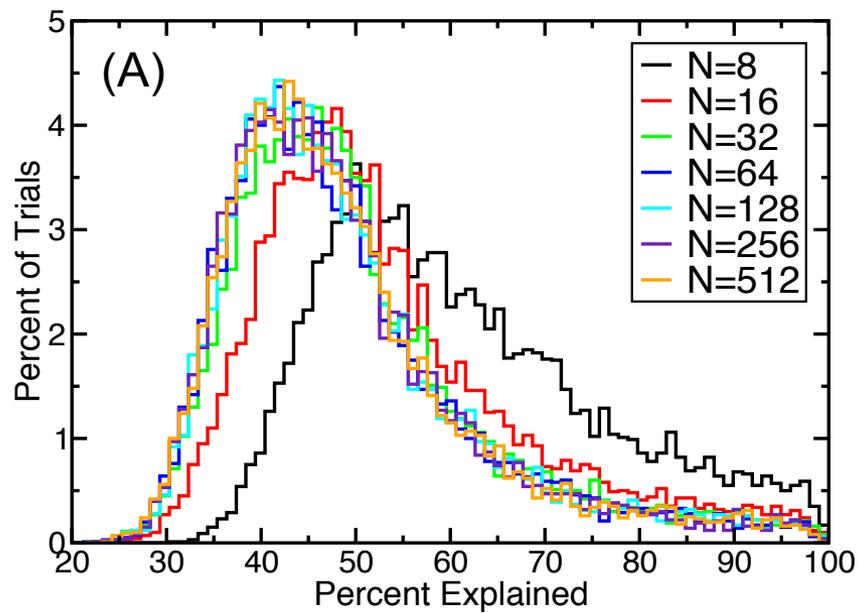
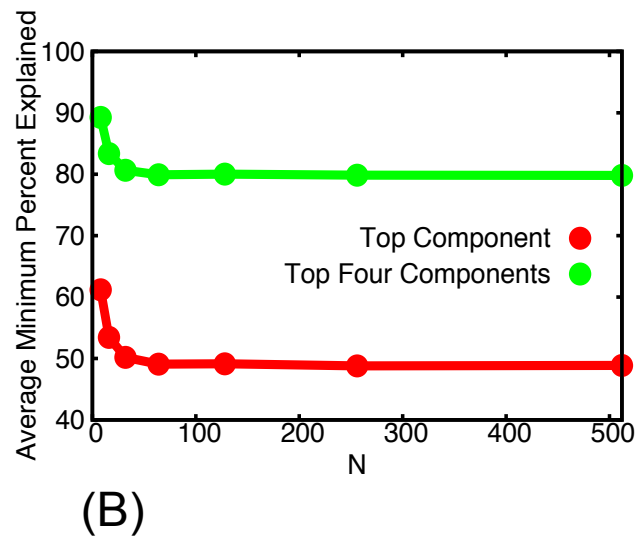
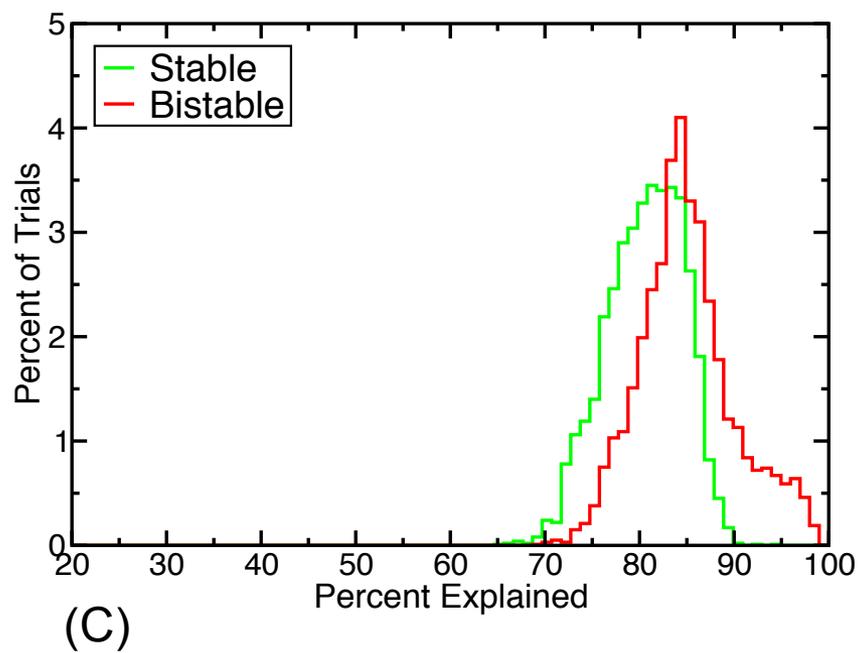
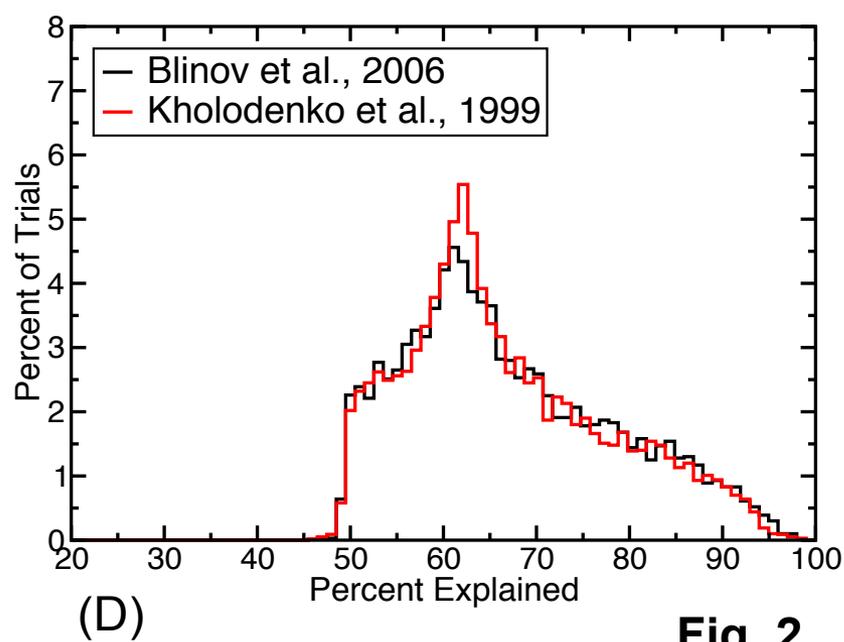

**Fig. 2**

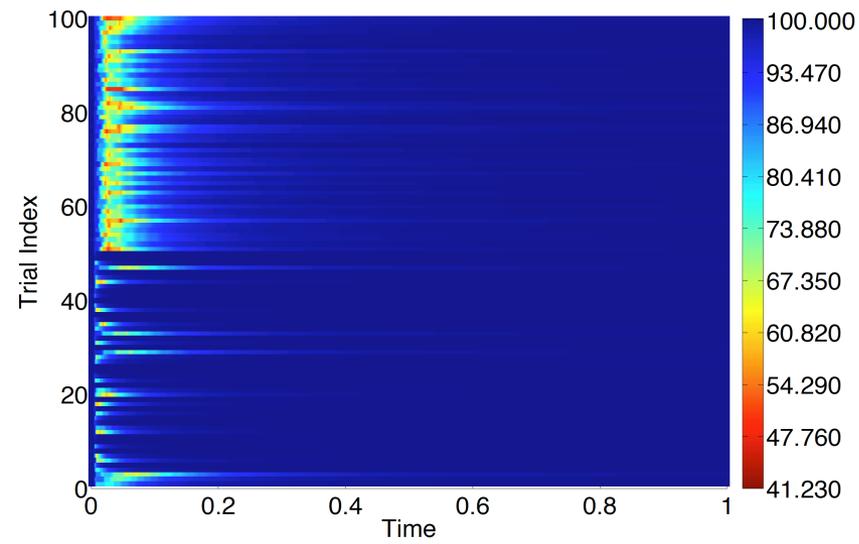 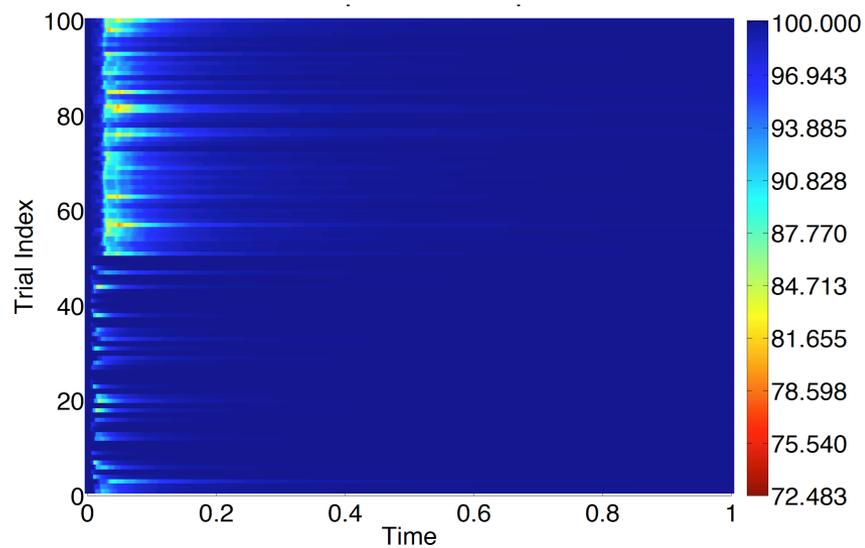 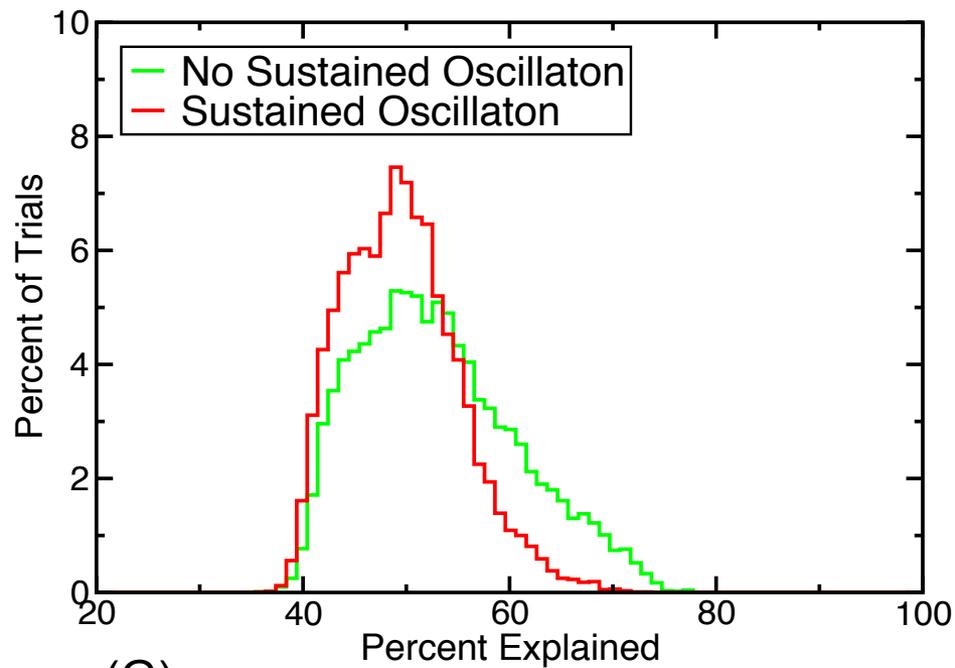

Fig. 2

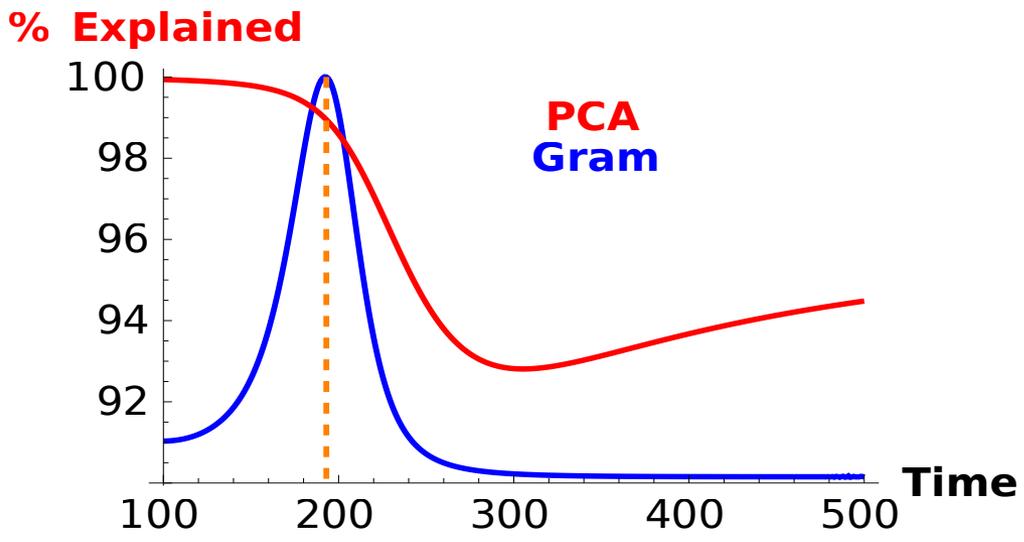 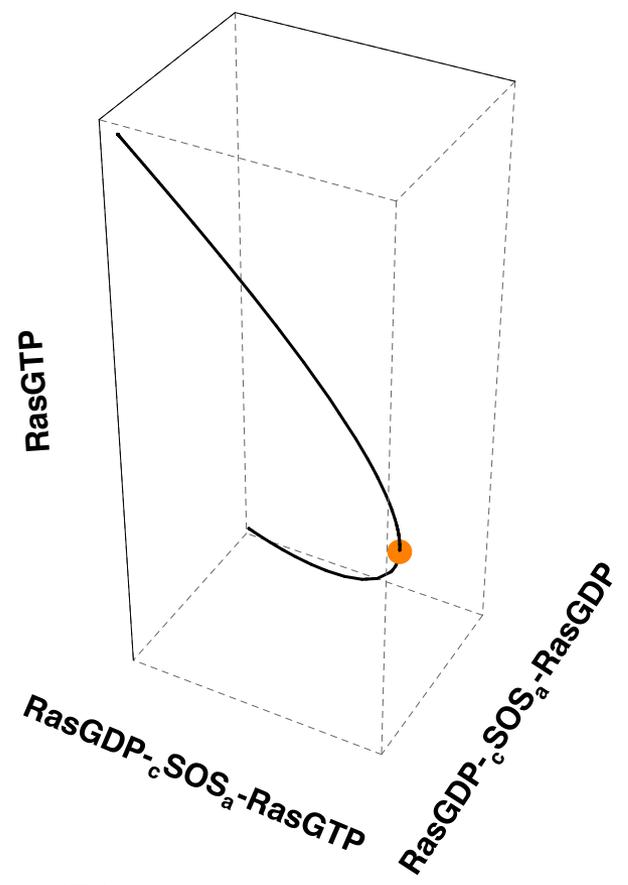

Fig. 3

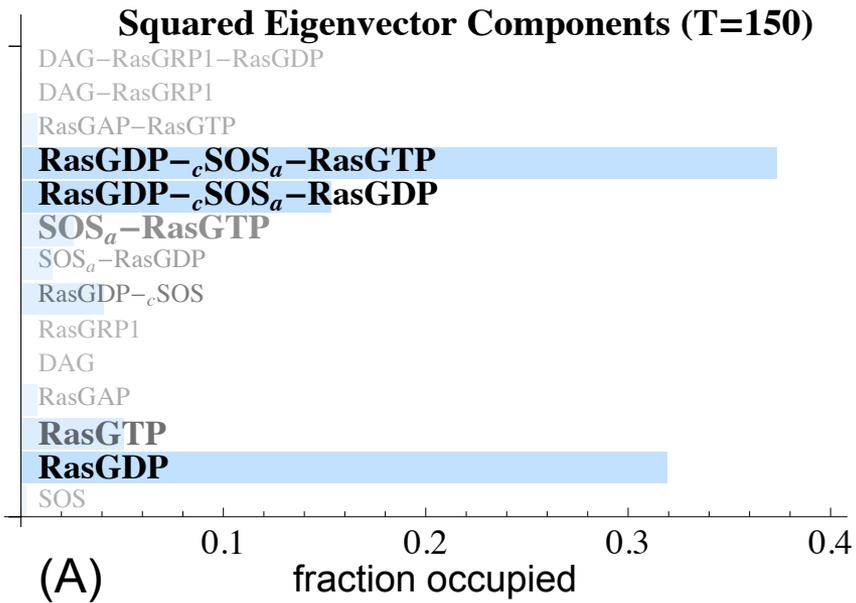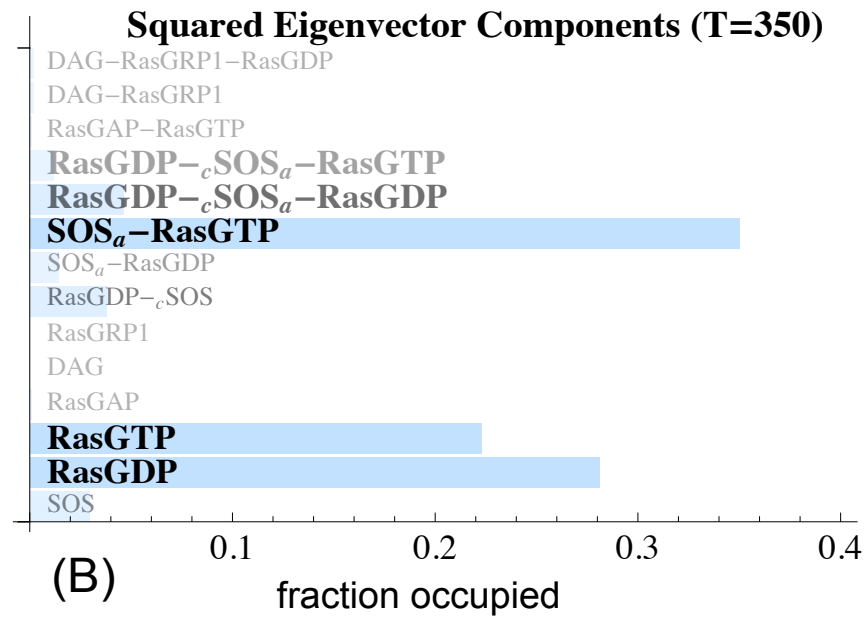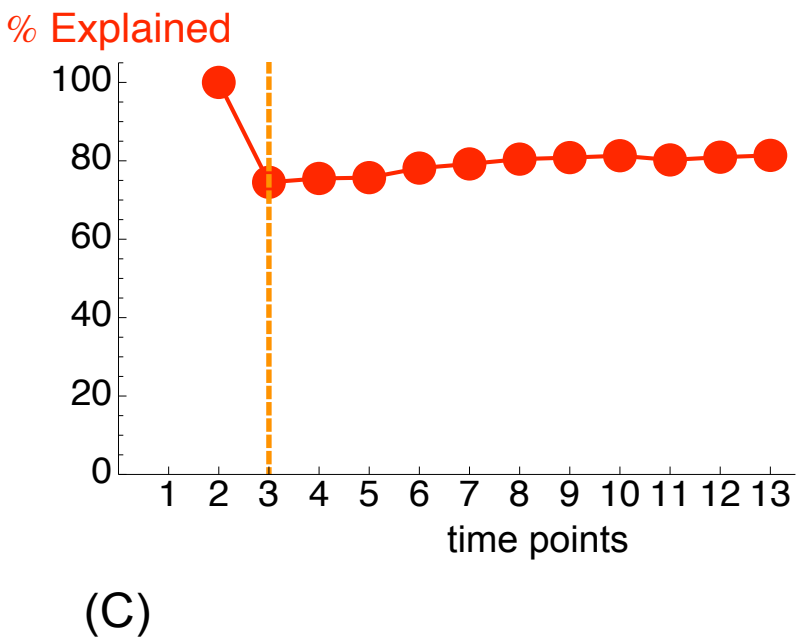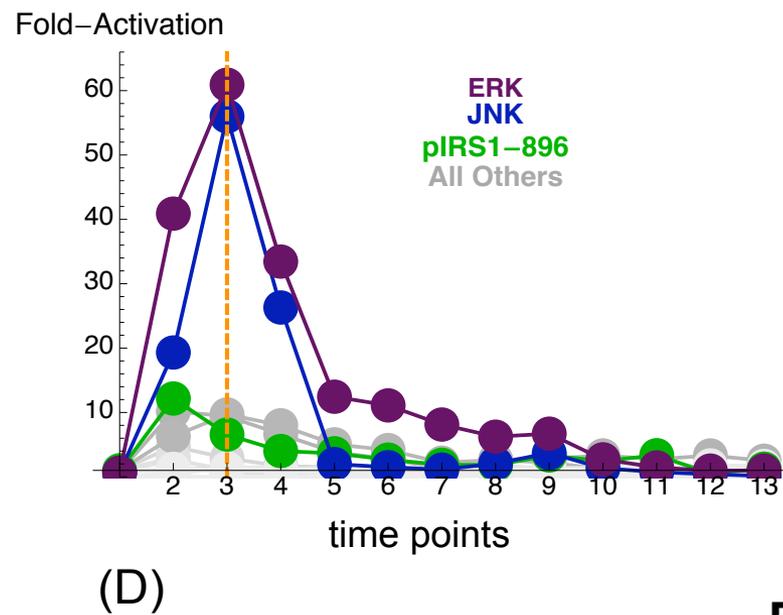

Fig. 4